\documentclass[aps, pra, article, amsmath, amssymb, graphicx, longbibliography,nofootinbib]{revtex4-1}

\usepackage{bm}
\usepackage{graphicx}
\usepackage{epsfig}
\usepackage{amsthm}
\usepackage{longtable}
\usepackage{booktabs}
\usepackage{epstopdf}
\usepackage[mathlines]{lineno}

\usepackage{grffile}
\usepackage{float}

\usepackage{amsthm}

\usepackage{pifont}

\begin{document}

\title{Gauge invariance of quantum electrodynamics \\of multi-electron atoms}

\author{M. I. Krivoruchenko}
\address{National Research Centre "Kurchatov Institute" \\ pl. Academika Kurchatova 1, 123182 Moscow, Russia}

\begin{abstract}
The proof of gauge invariance of the quantum electrodynamics of photons and electrons does not apply directly to the quantum electrodynamics of photons, electrons, and nuclei because multi-electron atoms belong to the space of asymptotic states of the extended theory. We offer two possible ways to circumvent this problem and prove, using a fairly general model for the description of nucleon-nucleon interaction in nuclei, the gauge invariance of the masses and electromagnetic form factors of multi-electron atoms in all orders of perturbation theory.
\end{abstract}

\maketitle


\section{Introduction}
\renewcommand{\theequation}{I.\arabic{equation}}
\setcounter{equation}{0}

Already in his first paper on the quantum electrodynamics of photons and electrons (QED), Feynman suggested the gauge invariance of the new theory,
based on the conservation of electromagnetic current \cite{Feynman1949}.
Specific arguments in support of his claim came later from
the Ward identity \cite {Ward1950} and generalizations
of the Ward identity derived by Green \cite {Green1953}, Fradkin \cite {Fradkin1955}, and Takahashi \cite {Takahashi1957}.
The Ward-Green-Fradkin-Takahashi (WGFT) identity makes it possible to rigorously
prove the gauge invariance of multi-loop diagrams in QED
after truncating and placing momenta of the external electron lines on the mass shell \cite{BjorkenDrell,BogoliubovShirkov1980},
provided the external electron lines do not contain self-energy insertions
\cite{Bialynicki1967}.
The electron self-energy operator turns out to be gauge dependent to one loop, as shown explicitly in Ref.~\cite{Itzykson1980}, and
to all orders of perturbation theory \cite{JohnsonZumino1959}.
A thorough analysis of the self-energy insertions carried out by Bialynicki-Birula \cite{Bialynicki1970}
revealed the gauge invariance of the renormalized cross sections in all orders of perturbation theory,
which essentially completed the proof of the gauge invariance of QED.

The non-covariant Coulomb gauge is popular
because it is suitable for explicitly solving the constraint field equations,
which allows passing from classical electrodynamics to QED with the use of canonical quantization. Equivalence of the
covariant Lorenz gauge and the non-covariant Coulomb gauge implies relativistic covariance of QED. Modeling
of electromagnetic interactions in multi-electron atoms usually involves the Lorenz and Coulomb gauges.

The static interaction potential of electrons in the Lorenz gauge is known as the Coulomb-Gaunt potential \cite{Gaunt1929}:
\begin{equation}
V_{\mathrm{CG}}(\mathbf{r}) =  \frac{e^2}{4\pi r}\left (1 - \mbox{\boldmath$\alpha$}_1 \cdot \mbox{\boldmath$\alpha$}_2\right).
\label{CGpot}
\end{equation}
Here, the matrices $\mbox {\boldmath$\alpha$}_{1,2}$ 
are the electron velocity operators divided by the speed of light $c$,
$\mathbf{r} = \mathbf{r}_1 - \mathbf{r}_2$, $\mathbf{r}_{1,2}$ are the electron coordinates, $r = |\mathbf{r}|$, and $e$ is the electron charge.
The potential $V_{\mathrm{CG}}(\mathbf{r})$ describes the lowest-order photon-exchange static interaction of two electrons.
The second term is responsible for the energy of the magnetostatic interaction; it is
of the same order $\sim v^2/c^2$ as the retardation correction to the photon exchange.
For the innermost shells $v/c \sim 1/r \sim \alpha Z$, where $ \alpha = e^2/(4\pi)$ and $Z$ is the nuclear charge.

The Coulomb-Breit static potential can be derived using the Coulomb gauge \cite{Breit1929}:
\begin{equation}
V_{\mathrm{CB}}(\mathbf{r}) = \frac{e^2}{4\pi r}\left(1 - \frac{\mbox{\boldmath$\alpha$}_1 \cdot \mbox{\boldmath$\alpha$}_2 + \mbox{\boldmath$\alpha$}_1 \cdot \mathbf{n} \mbox{\boldmath$\alpha$}_2 \cdot \mathbf{n}}{2}\right),
\label{CBpot}
\end{equation}
where $\mathbf{n} = \mathbf{r}/r$.
$V_{\mathrm{CB}}(\mathbf{r})$ is exact to order $v^2/c^2$; the retardation correction appears in order $ v^4/c^4 $.

These potentials are used to construct initial approximations in various problems of atomic physics. The gauge dependence of the results can be estimated, e.g., by comparing the contributions of the magnetostatic interaction energy to the two-electron ionization potentials (DEIPs). For noble gas atoms, the difference ranges from 0.02 eV in the $K$ shell of the neon atom to 56 eV in the $K$ shell of the ruthenium atom \cite{Niskanen2011}. In many applications, the accuracy of tens of electron volts is obviously insufficient.
In order to identify the resonant enhancement in a neutrinoless double-electron capture, the excitation energy of the electron shells should be determined with an accuracy of 10 eV or higher \cite{Krivoruchenko2011}. Such high accuracy is necessary to match the high accuracy of the Penning traps when measuring the atomic mass differences \cite{Blaum2006,74Se2}, as well as the natural widths of the excited electron shells \cite{Campbell2001}.

In the Lorenz and Coulomb gauges, the interaction energies of two electrons in the innermost shells coincide with an accuracy of $\alpha^4 Z^3$, provided that the Coulomb-Gaunt potential is supplemented by a retardation photon-exchange correction \cite{Blaum:2020}.
The inclusion of two-photon exchange together with the vacuum polarization, electron self-energy and QED counterterms
ensures the accuracy of calculating the energy spectrum $\sim \alpha^4 Z^2$ \cite{Shabaev:1993}.
In higher orders of QED perturbation theory, only isolated examples 
illustrate gauge invariance. In the modeling heavy atoms, where core electrons are essentially relativistic, the high orders play an important role. For example, in a lead atom, the $K$ shell electrons move at a speed of $v \sim 0.6 c$. The single-electron ionization potentials of heavy atoms are known with an accuracy of about 10 eV \cite{LARK}. Comparison of the DEIP extracted from Auger spectroscopy data with the results of calculations based on the software package G{\lowercase{\scshape RASP}}2K \cite{GRASP1989,Grant2007} shows that the uncertainties in DEIP are about 60 eV \cite{Blaum:2020}. The purpose of this paper is to find out whether it is possible to exclude any connection of uncertainties inherent in modeling atomic structure with the choice of gauge in all orders of QED perturbation theory.

The demonstration of gauge invariance is also crucial for determining fundamental constants and testing QED with light atoms in high orders of perturbation theory \cite{Karshenboim:2005}.

The proof of the gauge invariance of QED relies on the fact that
electron lines  of the Feynman diagrams either form closed loops or extend to infinity.
In scattering theory, particles that propagate to infinity form a Fock space of asymptotic states.
Photons and electrons (and positrons) are the particles that create asymptotic states within QED.
It is worth noting that positronium is a resonance in the $2\gamma$, $3\gamma$, and other multi-photon channels;
it annihilates, hence
does not propagate to infinity, and therefore does not belong to the space of asymptotic states.
In contrast, the quantum electrodynamics of photons, electrons, and nuclei predicts bound states
such as a hydrogen atom; when scattered, bound states either disintegrate or propagate to infinity.
Acting on the vacuum, bound states generate a new set of asymptotic states, thereby expanding the Fock
space of asymptotic states of elementary particles.
The existence of bound states takes us beyond the assumptions in which the standard proof of QED's gauge invariance is performed.

In this paper, we develop two arguments proving the gauge invariance of the quantum electrodynamics of photons, electrons,
and nuclei. One argument is based on the analyticity of scattering amplitudes. The other is based
on the WGFT identity generalized to electron and proton loops with vertices of photons,
strongly interacting mesons, and an  auxiliary fictitious scalar particle.
In both cases, the problem of bound states as
asymptotic states of the extended theory
can be successfully circumvented.

\section{Analytical properties of scattering amplitude}
\renewcommand{\theequation}{II.\arabic{equation}}
\setcounter{equation}{0}

The renormalized QED amplitudes depend on the kinematic invariants
\[
s_{ij} = (k_i + k_j)^2,\;\;\;s_{ijk} = (k_i + k_j + k_l)^2,\;\;\; \ldots,
\]
where $k_i$ are the incoming and outgoing electron and photon four-momenta. The renormalized mass $m$ and charge $e$ enter the amplitudes as parameters. In what follows, the renormalized amplitudes are denoted by $\mathcal{A} (s_{ij},\ldots)$, and the helicity
indices are suppressed. $\mathcal{A} (s_{ij},\ldots)$ is gauge-invariant provided its functional form and the parameters $m$ and $e$
do not vary under the gauge transformations. It is sufficient to show that $\mathcal{A} (s_{ij},\ldots)$ remains unchanged under variations of longitudinal components of photon propagator and photon polarization vectors.

The scattering amplitudes are known to be boundary values of the analytical functions of the kinematic invariants \cite{Eden1966}. By crossing symmetry, the scattering amplitudes in different channels are expressed in terms of a unique analytic function at different kinematic invariants' values. Analyticity is one of the postulates of axiomatic quantum field theories.

Singular points of Feynman diagrams are determined by renormalized masses of particles entering the Lagrangian. Summation of perturbation series can generate additional singularities, including poles corresponding to bound states. Since each member of the perturbation series depends on renormalized masses, additional singular points are also determined by renormalized masses.

The renormalized amplitudes of QED are gauge-invariant,
so the renormalized electron mass and charge are gauge-invariant.
As a consequence, the mass counter-term $\delta m$ and the renormalization constant $Z_3$ of the photon propagator
emerge as gauge-invariant values regardless of the regularization setup
because of the relations $m = m_0 + \delta m$ and $e = \sqrt{Z_3}e_0$ and the obvious fact that the bare electron mass $m_0$ and charge $e_0$
do not depend on the gauge.
These conclusions do not extend to the renormalization constants $Z_1$ and $Z_2$,
which are gauge-dependent in general \cite{Itzykson1980,JohnsonZumino1959}. The Ward identity implies $Z_1 = Z_2$.

The analytical continuation of the two-body Green's function in energy below the elastic threshold is used to demonstrate the gauge invariance of the positronium binding energy in Ref.~\cite{Feldman:1980}.
Given that positronium appears as a Breit-Wigner pole in the physical region of scattering of two, three, or more photons, it is not obligatory to use analytical continuation in this context.
The standard proof techniques in QED \cite{BjorkenDrell,BogoliubovShirkov1980,Bialynicki1967,Bialynicki1970} 
are perfectly suited for the annihilation channels.

We start by considering the quantum electrodynamics of photons, electrons, and protons (QED+).
In a wide range of atomic spectroscopy problems, a proton can be considered as an elementary
charged spin-1/2 fermion. This idealization allows the problem to be formulated accurately and
the proof to be carried out using the standard QED techniques.

\subsection{Forward scattering amplitude}

\begin{figure} [t] %
\begin{center}
\includegraphics[angle = 0,width=0.382\textwidth]{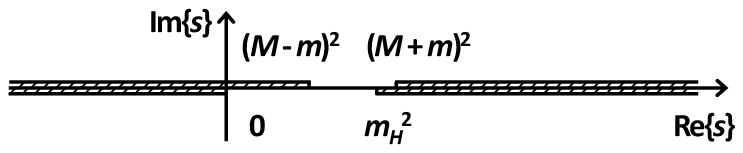}
\caption{
Analytical structure of the forward scattering amplitude $e^- + p \to e^- + p$ in the complex $s$-plane.
The branch cut $(M + m)^2 < s < +\infty$ is associated to the elastic $ e^- +p \to  e^-+ p$ scattering of the $s$-channel.
The branch cut $m_H^2 < s < +\infty$, which starts with the ground-state mass of hydrogen atom, $m_H$,
occurs after the series summation of perturbation theory.
The branch cut $-\infty < s < (M-m)^2$ is associated to the elastic $ e^- +\bar{p} \to  e^- + \bar{p}$ scattering
of the $u$-channel. The branch cut $-\infty < s < 0$ is associated to the $e^-+e^+ \to p + \bar{p}$ scattering
of the $t$-channel.
}
\label{figt0}
\end{center}
\end{figure}

In the $e^- + p$ channel, there is a single bound state, hydrogen atom $H$, and an infinite number of semi-bound states, $H^*$, which are excited states of $H$. The excited states decay with the emission of photons: $H^* \to H + \gamma + \ldots$. The forward $ e^- + p \to e^- + p$ scattering amplitude, $\mathcal{A}(s,t = 0)$, has a pole in $s \equiv s_{12} = (k_1 + k_2)^2$ at the ground-state mass $ m_H^2 < {s_0} \equiv (M + m)^2 $, with $k_1$ and $k_2$ being four-momenta of the incoming electron and proton, respectively; $s_0$ is the elastic threshold and $M$ is the proton mass. The variable $t$ equals $t= (k_3 - k_1)^2$,
where is $k_3$ the four-momentum of outgoing electron. The unitary cut in the variable $s$ extends to the right half-axis  $m_H^2 < s < +\infty$.
The left branch cut, associated with the crossing $u$-channel $e^- + \bar{p} \to  e^- + \bar{p}$, extends to $s \in (- \infty,(M - m)^2)$. The $t$-channel $e^- + e^+ \to  p + \bar{p}$ contributes to the discontinuity
for $s \in (- \infty,0)$. The discontinuity across the interval $s \in (m_H^2,s_0)$ is associated with the excited states' radiative decays. The analytical properties of $\mathcal{A}(s,t = 0)$ on the physical sheet of the Riemann surface are shown in Fig.~\ref{figt0}. They are determined by the renormalized masses of electron, proton and the ground-state mass of hydrogen atom. The poles of $\mathcal{A}(s,t = 0)$ corresponding to the semi-bound states are shifted down from the cut $(m_H^2,s_0)$ to the non-physical sheet adjacent to the physical sheet of the Riemann surface. The problem is to prove that the pole positions are independent of the gauge. We also show that electromagnetic diagonal and transition form factors of hydrogen are gauge-invariant.

Double spectral representation details the relationship between the parameters of a hydrogen atom and the analytical properties of the $e^- + p \to e^- + p$ scattering amplitude.

\subsection{Double spectral representation}

Two-body scattering amplitudes admit double spectral representations \cite{Mandelstam:1958,Mandelstam:1959,Eden1966}.
The spectral forms depend on the renormalized masses of particles that generate the space of asymptotic states,
including bound states when they exist.

\begin{figure} [!t] %
\begin{center}
\includegraphics[angle = -90,width=0.5\textwidth]{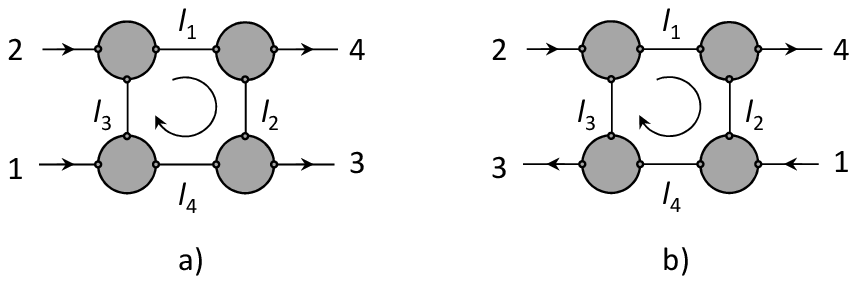}
\caption{
Diagrams to determine boundary of the regions of spectral support.
Arrows indicate momentum flow.
}
\label{fig8}
\end{center}
\end{figure}

Figure~\ref{fig8}~(a,b) shows diagrams to determine the boundary curves of the
double spectral forms of two-body  scattering amplitudes in the $1+2 \to 3+4$ process.
The masses of particles in the loops are denoted by $m_{ij}$ with $i,j=1\ldots4$.
A particle with the mass $m_{ij}$ has common vertices with two internal lines and two external lines $i$ and $j$.
A one-to-one correspondence exists between the external lines and the vertices. The external line 1 is attached to the vertex 1, etc.
The external particle momenta $k_i$ satisfy the energy-momentum conservation $k_1 + k_2 = k_3 + k_4$.
The Mandelstam variables equal $s = (k_1 + k_2)^2$, $t = (k_4 - k_2)^2$ and $u = (k_3 - k_2)^2$.
The diagrams a) and b) go over each other after the substitutions
$1 \leftrightarrow 3$, $k_1 \leftrightarrow - k_3$, $s \leftrightarrow u$.

Let $q$ be the internal particle momentum in the loop, over which the integration is performed.
The  internal particle momenta shown in Figs.~\ref{fig8}~(a,b) equal
\begin{equation} \label{cccc}
\begin{array}{llll}
l_1 = q, & l_2 = q - k_4, & l_3 = q - k_2, & l_4 =
\left\{
\begin{array}{ll}
q - k_4 - k_3, & ~~~~\mathrm{Fig.~\ref{fig8}~a)},\\
q - k_2 + k_3, & ~~~~\mathrm{Fig.~\ref{fig8}~b)}.
\end{array}%
\right.
\end{array}%
\end{equation}%
The energy-momentum is conserved: $l_1 - l_3 = k_2$, etc.
At the boundary of double spectral form
$l_i$ satisfy the on-shell conditions:
\begin{equation} \label{bbbb}
\begin{array}{lllll}
l_1^2 = m_{24}^2,& l_4^2 = m_{13}^2, & l_2^2 = m_{34}^2, & l_3^2 = m_{12}^2, & ~~~~\mathrm{Fig.~\ref{fig8}~a)}, \\
l_1^2 = m_{24}^2,& l_4^2 = m_{13}^2, & l_2^2 = m_{14}^2, & l_3^2 = m_{23}^2, & ~~~~\mathrm{Fig.~\ref{fig8}~b)}.
\end{array}%
\end{equation}%
A non-vanishing double spectral form exists for
\begin{equation} \label{aaaa}
\begin{array}{lll}
s > (m_{24} + m_{13})^2,& t>(m_{12} + m_{34})^2, & ~~~~\mathrm{Fig.~\ref{fig8}~a)}, \\
u > (m_{24} + m_{13})^2,& t>(m_{14} + m_{23})^2, & ~~~~\mathrm{Fig.~\ref{fig8}~b)}.
\end{array}%
\end{equation}%

The boundary of regions of spectral support 
can be parameterized in terms of the scattering angles \cite{Mandelstam:1958}.
The bounding curves are determined by requiring that the determinant of the Gram matrix $\Delta(s,t) = \det||l_i l_j||$ be equal to zero:
\begin{equation} \label{det}
\Delta(s,t) = 0.
\end{equation}
The explicit form of $\Delta(s,t)$ is given by Kibble \cite{Kibble:1960}
(see also \cite{Itzykson1980}).
$\Delta(s,t)$ of Fig. 2 a) turns to $\Delta(s,t)$ of Fig. 2 b) upon the substitutions $s \leftrightarrow u$,
$m_1 \leftrightarrow m_3$, $m_{14} \leftrightarrow m_{34}$ and $m_{12} \leftrightarrow m_{23}$.\footnote{\footnotesize{
Mandelstam \cite{Mandelstam:1959} gives the equations of bounding curves for a number of the
pion-nucleon scattering diagrams with low-mass intermediate states. For diagram e) in Fig. 5 of his paper, Eq.~(\ref{det})
gives
\begin{equation} \label{lineline}
(u-(M+2\mu )^{2})(u-(M-2\mu )^{2})(s-(M+\mu )^{2})(s-(M-\mu )^{2})-16\mu
^{2}M^{2}su-16\mu ^{2}(M^{2}-\mu ^{2})^{2}(t-M^{2})=0, \tag{$^*$}
\end{equation}
where $\mu $ is the pion mass, and $M$ is the nucleon mass. This equation
is different from Eq.~(4.3a) of Ref.~\cite{Mandelstam:1959}. The
region of spectral support of diagram f) in Fig.~5 can be obtained from Eq.~(\ref{lineline}) 
using the crossing symmetry $s\leftrightarrow u$, which permutes the diagrams e) and
f). Shirkov with co-workers \cite{Shirkov:1969} quote the equations of
bounding curves of Ref.~\cite{Mandelstam:1959} and discuss an additional
diagram a) in Fig.~47, for which we obtain
\begin{equation*} \label{line}
(t - 4M^{2})(s-(M+\mu )^{2})(s-(M-\mu )^{2})-4M^{4}s+4(M^{2}-\mu
^{2})^{2}(s-2\mu ^{2})
=0. 
\end{equation*}
This equation is different from Eq.~(23.12) of Ref.~\cite{Shirkov:1969}.
The discrepancies are related to erroneous identifications of scattering angles
in the intermediate state of the diagrams.
The equation preceding Eq.~(23.12), 
e.g., should look like
\begin{equation*} \label{linelineline}
z_2 = z_3 = \left. \frac{t}{2q^2} + 1\right|_{u = M^2}.
\end{equation*}
}}

\begin{figure} [!t] %
\begin{center}
\includegraphics[angle = 0,width=0.445\textwidth]{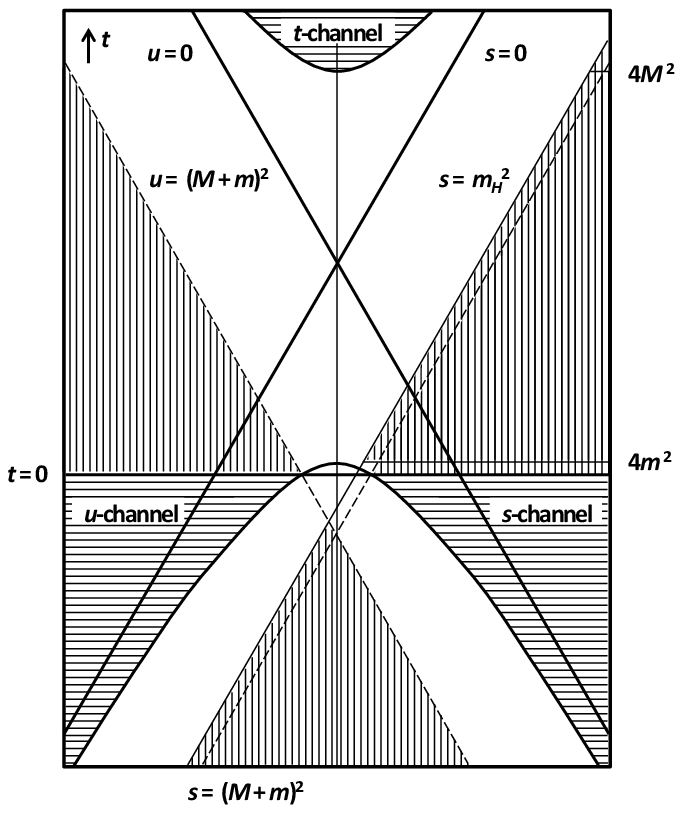}
\caption{
Mandelstam diagram of the $e^- + p \to e^- + p$ scattering.
$m$ is the electron mass, $M$ is the proton mass, $m_H$ is hydrogen mass in the ground state.
The spectral support regions are indicated by vertical hatching. Physical regions are marked with horizontal hatching.
}
\label{fig9}
\end{center}
\end{figure}

Consider the $e^- + p \to e^- + p$ scattering.
The lines 1 of Fig.~\ref{fig8}~(a,b)
denote the proton   with the mass $m_1 = M$ and
the lines 2 denote the electron with the mass $m_2=m$.
The intermediate states of the interest are as follows:

1.~Diagram a). The lines with momenta $l_4$ and $k_3$ describe the proton: $m_{13} = m_3 = M$.
The lines with momenta $l_1$ and $k_4$ describe the electron: $m_{24} = m_4 = m$.
The lines with momenta $l_2$ and $l_3$ describe the massless photons: $m_{12}=m_{34} =0$.
Constraints (\ref{aaaa}) imply $s \geq (M + m)^2$ and $t \geq 0$. Equation~(\ref{det}) gives
\begin{eqnarray}
&&t^2(s - (M + m)^2)(s - (M - m)^2) = 0. \label{ep1}
\end{eqnarray}
The boundary of the support consists of the half-lines $s = (M + m)^2 \cap t \geq 0$ and $s \geq (M + m)^2 \cap t = 0$.
The region of the non-vanishing double spectral form is localized in the upper right part of the Mandelstam diagram, Fig.~\ref{fig9}.
The double spectral forms are shown without the exact adherence the physical proportions between $M$, $m$, and $m_H$.
The asymmetry of the spectral support upon the $s \leftrightarrow u$ transformation is determined by hydrogen atom.
Singular points of the forward scattering amplitude in the complex $s$ plane, attached to the $t = 0$ axis of the diagram,
are shown in Fig.~\ref{figt0}.

2.~Diagram a). The external line 4 is the electron, $m_4 = m$,
and the external line 3 is the proton, $m_3 = M$.
The line 13 with a momentum $l_4$ describes hydrogen atom, $m_{13} = m_H$.
The line 24 with a momentum $l_1$ is the photon, $m_{24}=0$. The lines 12 and 34 with momenta $l_3$ and $l_2$ are the electrons, $m_{12}=m_{34} = m$.
Constraints (\ref{aaaa}) imply $s \geq m_{H}^2$ and $t \geq 4m^2$.
Equation~(\ref{det}) gives
\begin{eqnarray}
&&t(t-4m^2)(s - m_H^2)^2  = 0. \label{ep2}
\end{eqnarray}
The boundary is determined by the half-lines $s = m_H^2 \cap t \geq 4m^2 $ and $s \geq m_{H}^2 \cap t = 4m^2$.
The corresponding region
has the shape of triangle shifted in relation to the case considered above (see Fig.~\ref{fig9}).

3.~Diagram a). The line 13 is the photon, $m_{13} = 0$. The line 2-4 is hydrogen atom, $m_{24}=m_H$. The lines 12 and 34 are the protons, $m_{12}=m_{34} =M$.
The external line 4 is the electron, $m_4 = m$, and the external line 3 is the proton, $m_3 = M$.
Constraints (\ref{aaaa}) imply $s \geq m_{H}^2$ and $t \geq 4M^2$. Equation~(\ref{det}) gives
\begin{eqnarray} \label{ep3}
t(t-4M^2)(s - m_H^2)^2  = 0.
\end{eqnarray}
The boundary is determined by the half-lines $s = m_H^2 \cap t \geq 4M^2$ and $s \geq m_{H}^2 \cap t = 4M^2$.

The considered diagrams give contributions to the double spectral form in the region of positive $s$ and $t$.

4.~Diagram a). The line 13 describes hydrogen atom, $m_{13} = m_H$. The line 24 is the photon, $m_{24}=0$.
The line 1-2 is the electron, $m_{12} = m$, the line 34 and the external line 4 describe the proton, $m_{34} = m_4 = M$,
and the external line 3 is the electron, $m_3 = m$.
In comparison with the previous diagrams, we interchange the electron and proton lines in the final state.
Constraints (\ref{aaaa}) imply $s \geq m_{H}^2$ and $u \geq (M + m)^2$.
The boundary equation reads:
\begin{eqnarray} \label{ep4}
(s - m_H^2)^2 (u - (M - m)^2) (u - (M + m)^2) = 0.
\end{eqnarray}
The boundary is determined by the half-lines $s = m_H^2 \cap u \geq (M + m)^2$ and $s \geq m_{H}^2 \cap u =  (M + m)^2 $.
The corresponding region occupies the lower part of the Mandelstam diagram in Fig.~\ref{fig9}.

5.~Diagram a). The line 13 is the photon with the mass $m_{13} = 0$. The line 24 is hydrogen atom with the mass $m_{24}=m_H$. The line 12 is the proton, $m_{12}=M$, the line 34 is the electron, $m_{34} = m$. The line 4 is the proton, $m_4 = M$,
and the line 3 is the electron with the mass $m_3 = m$.
Constraints (\ref{aaaa}) imply $s \geq m_{H}^2$ and $u \geq (M + m)^2$.
Equation~(\ref{det}) gives
\begin{eqnarray} \label{ep5}
(s - m_H^2)^2 (u - (M - m)^2) (u - (M + m)^2) = 0.
\end{eqnarray}
The boundary is made up of the half-lines $s = m_H^2 \cap u \geq (M + m)^2$ and $s \geq m_{H}^2 \cap u =  (M + m)^2$.
The corresponding region is the same as in the previous case.

The last two diagrams give contributions to the double spectral form in the region of positive $s$ and $u$.

6.~Diagram b). The line 13 is the proton, $m_{13} = M$. The line 24 is the electron, $m_{24}=m$. The lines 12 and 34 are the photons, $m_{12}=m_{34} = 0$.
The external line 4 is the electron, $m_4 = m$,
and the external line 3 is the proton, $m_3 = M$.
Constraints (\ref{bbbb}) imply $u \geq (M + m)^2$ and $t \geq 0$.
Equation~(\ref{det}) gives
\begin{eqnarray} \label{ep1a}
t^2(u - (M + m)^2)(u - (M - m)^2) &=& 0.
\end{eqnarray}
The boundary is made up of the half-lines $u = (M + m)^2 \cap t \geq 0$ and $u \geq (M + m)^2 \cap t = 0$.
The diagram contributes to the double spectral form in the left upper part of the Mandelstam diagram.

7.~The intermediate state $H$ in the $e^- + p \to e^- + p$ scattering generates in the amplitude a pole at
\begin{eqnarray} \label{epH}
s = m_H^2,
\end{eqnarray}
In Fig.~\ref{fig9}, this pole is shown as the solid line parallel to the dashed line of the $e^- + p$ elastic threshold.

The considered list of the diagrams is restricted to the low-mass intermediate states.
The intermediate states with higher masses correspond to regions that are subsets of the constructed regions
of the non-vanishing double spectral form.

Summing up, the boundary of the support of the double spectral form, hence the scattering amplitude,
depends on the bound state mass $m_H$. 

\begin{figure} [!b] %
\begin{center}
\includegraphics[angle = -90,width=0.39\textwidth]{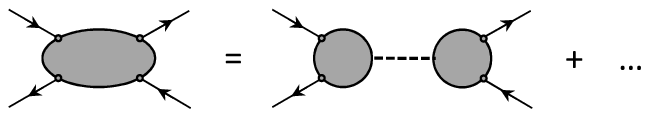}
\caption{
Two-fermion Green's function continued analytically to the complex $s$-plane
and expanded in the Laurent series in a neighborhood of two-fermion bound state of the mass $m_H$.
Dashed line denotes the bound-state propagator $i/(s - m_H^2)$.
Gray-shaded circles denote the Bethe-Salpeter wave functions.
Dots stand for less-singular parts of the diagram.
}
\label{figgauge1}
\end{center}
\end{figure}

\section{Gauge invariance of masses and electromagnetic form factors}
\renewcommand{\theequation}{III.\arabic{equation}}
\setcounter{equation}{0}

In the previous section, we described the relationship between the ground-state mass of hydrogen atom and the singularities of the $e^- + p \to e^- + p$ scattering amplitude. Hydrogen atom is not among the asymptotic states of the $e^- + p \to e^- + p$ channel. Therefore, the standard proofs of the gauge invariance of QED apply to all Feynman diagrams of the elastic channel at any fixed order of perturbation theory. The sum of the series, each term of which is gauge-invariant, determines the gauge-invariant quantity.%
\footnote{\footnotesize{
It is possible to overcome difficulties associated with the asymptotic character of perturbation series in QED-like theories \cite{Dyson:1952}
by summing infinite sets of diagrams using equation-solving techniques.
The Bethe-Salpeter equation \cite{Itzykson1980,Nakanishi:1969} is used to add up diagrams for the two-particle Green's function, for example.
The $e^-p \to e^-p$ Green's function is well known in non-relativistic theory and is utilized in a variety of applications.
Divergent series are frequently calculable using the Borel summation method \cite{Shawyer:1994}. Ref.~\cite{Krivoruchenko:2006} contains a physical example of chiral
symmetry in relation to the summing of asymptotic series. The invariance of each term of the series is regarded a sufficient
condition for the invariance of the entire sum in the following, subject to the unique requirement that the entire sum be
calculated unambiguously.
}}
Upon the summation, the perturbation series of QED+ describe the bound state, hydrogen atom, and concurrently the associated singularities of the scattering amplitudes, which depend on $m_H$. These observations indicate the gauge invariance of the mass of the hydrogen atom. We detail this argument below and generalize it to the ground and excited states and electromagnetic form factors of a hydrogen atom and multi-electron atoms.

\subsection{Hydrogen atom}

\textit{Masses of ground and excited states.} Techniques developed to prove the gauge invariance of QED apply
to diagrams of QED+, electron, and proton lines either form closed loops or belong to asymptotic states. At any fixed order of perturbation theory, the forward scattering amplitude $\mathcal{A}(s,t = 0)$ of the process
\begin{equation}
e^-+p \to e^-+p
\end{equation}
is therefore gauge-invariant above the elastic threshold $s_0 = (m+M)^2$.
After summing up the perturbation series, the scattering amplitude remains gauge-invariant. As stated earlier, the scattering amplitudes are analytic functions of kinematic invariants. We continue $\mathcal{A}(s,t = 0)$ analytically from the upper edge of the unitary cut $(s_0,+\infty)$
to the complex $s$-plane. In the process of analytical continuation,
one finds a pole at $s = m_H^2$, corresponding to the ground-state hydrogen atom, and poles under the cut $(m_H^2,s_0)$, corresponding to semi-bound states of a hydrogen atom. A decomposition of two-body Green's function in a neighborhood of $s = m_H^2$ is shown in Fig.~\ref{figgauge1}. The ground-state mass, $m_H$, of hydrogen atom and the complex masses of semi-bound states, $m_H^*$, are gauge-invariant, since the amplitude $\mathcal{A}(s,t = 0)$ is gauge-invariant for $s \in (s_0,+\infty)$.
The analytical continuation of a gauge-invariant amplitude defined by the
perturbation series for $s \in (s_0,+\infty)$ produces a gauge-invariant amplitude
in the whole complex $s$-plane.
\begin{figure} [t] %
\begin{center}
\includegraphics[angle = 0,width=0.62\textwidth]{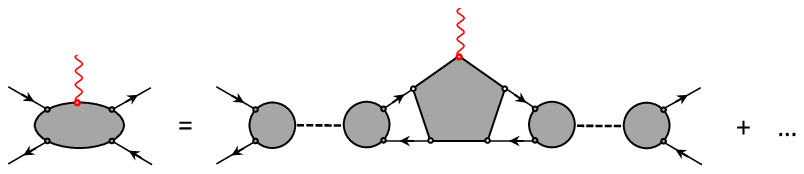}
\caption{
Five-point Green's function continued analytically in the invariant masses
of incoming and outgoing fermions to neighborhoods of poles
corresponding bound or semi-bound states. Solid lines identify electron and proton, and wavy lines represent the virtual photon. Arrows show electric charge flow. Dashed lines denote propagators of two-fermion bound or semi-bound states, gray-shaded circles denote their Bethe-Salpeter wave functions, and grey-shaded polygon with the virtual photon defines the electromagnetic form factor. Dots stand for less-singular parts of the diagram.
}
\label{figgauge2}
\end{center}
\end{figure}

It should also be pointed out that the photon propagator's regularization with an infrared photon mass
allows bypass infrared divergences in high orders of perturbation theory. Such a regularization scheme leaves
the fermion vertices transverse on the photon momentum. While a massive photon violates the gauge invariance of
the QED Lagrangian explicitly, the WGFT identity holds. When the proof of the gauge invariance is completed,
the photon mass can be set to zero (see, e.g., \cite{Bialynicki1970}).

\textit{Electromagnetic form factors.}
Let us consider the electron-proton scattering accompanied by a photon's emission, as shown in Fig.~\ref{figgauge2}. Diagram of Fig.~\ref{figgauge2} can be a part of a more general diagram. The gauge invariance of the electromagnetic form factor follows from the decomposition of the scattering amplitude in elementary blocks.
The central vertex shown by the grey-shaded polygon determines the electromagnetic form factor. We perform analytical continuation of the five-point Green's function to the complex plane of the invariant mass $s_i = P^2_i$ of the incoming $e^- + p$ system and then to the complex plane of the invariant mass $s_f = P^2_f$ of the outgoing $e^- + p$ system. In neighborhoods of the poles corresponding to the ground state of a hydrogen atom, the scattering amplitude $\mathcal{A} (s_f,s_i,\ldots)$ is factorized as shown in Fig.~\ref{figgauge2}. The dominant part of the diagram is proportional to $1/(s_i - m_H^2)/(s_f - m_H^{2})$. The next-to-leading terms of the expansion are of the order $O(1/(s_f - m_H^2))$ + $O(1/(s_i - m_H^2))$. The scattering amplitude constructed using the Green's function of Fig.~\ref{figgauge2} is gauge-invariant; thus, the electromagnetic form factor that appears in the leading order of the Laurent series also does not depend on the gauge.

The gauge invariance of the electromagnetic transition form factors can be demonstrated similarly. The relevant blocks of the amplitude
\begin{equation}
e^- + p \to \gamma^* + e^- + p
\end{equation}
are determined by poles in the variables $s_i$ and $s_f$ on the Riemann surfaces' unphysical sheets adjacent to the physical sheets along the cuts $(m_H^2,s_0)$. The photon $\gamma^*$ can be real, while masses of the incoming and outgoing semi-bound states are complex and different in general. The imaginary part of the masses determines the decay width of the semi-bound states.

\subsection{Multi-electron atoms}

\textit{Masses of ground and excited states.}
Multi-electron atoms are treated in the framework of quantum electrodynamics of photons, electrons, and nuclei. Nuclei are considered as elementary spin-1/2 fermions with the mass $ m_{\mathrm{nucl}} $ and charge $ - eZ > 0$. The proof proceeds by the analytical continuation of the forward scattering amplitude $\mathcal{A} (s_{ij},\ldots)$ of $Z$ electrons and a nucleus:
\begin{equation}
Ze +\mathrm{nucleus} \rightarrow Ze + \mathrm{nucleus}.
\end{equation}

Asymptotic states of this particular channel do not contain bound states.
Above the threshold $s_0 = (Z m + m_{\mathrm{nucl}})^2$
$\mathcal{A} (s_{ij},\ldots)$ is therefore gauge-invariant. Analytical continuation of $\mathcal{A} (s_{ij},\ldots)$ in the invariant mass $s \equiv P_{\mathrm{total}}^2$ of the initial state, assuming the other kinematic variables are fixed, leads to a pole at $s = m_{{\mathrm{atom}}}^2 < s_0$ corresponding to the ground state of the atom. Since $\mathcal{A} (s,\ldots)$ does not depend on the gauge, $m_{\mathrm{atom}}$ is also independent of the gauge. The same arguments apply to semi-bound atomic states localized below the cut $(m_{\mathrm{atom}}^2,s_0)$.

\textit{Electromagnetic form factors.}
Instead of the forward scattering, we consider the process
\begin{equation} \label{Zenucl}
Ze +\mathrm{nucleus} \rightarrow \gamma^* + Ze + \mathrm{nucleus},
\end{equation}
in which the state of the outgoing electrons and the nucleus is obtained applying a bust transformation to the state of the incoming electrons and the nucleus.
The form factor is identified as a multiplier of $1/(s_i-m_{\mathrm{atom}}^{*2})/(s_f-m_{\mathrm{atom}}^{**2})$ in the decomposition of the amplitude,
where $s_i = P_{i,\mathrm{total}}^2$ and $s_f = P_{f,\mathrm{total}}^2$ are the invariant masses of the incoming and outgoing particles excluding the photon $\gamma^*$
and $m_{\mathrm{atom}}^{*}$ and $m_{\mathrm{atom}}^{**}$ is the atomic mass of the ground state or an excited state.
The diagonal form factors are described with $m_{\mathrm{atom}}^{*} = m_{\mathrm{atom}}^{**}$. The gauge invariance
of the amplitude of the process (\ref{Zenucl}) implies the gauge invariance of the form factors.

\begin{figure} [!t] %
\begin{center}
\includegraphics[angle = 270,width=0.498\textwidth]{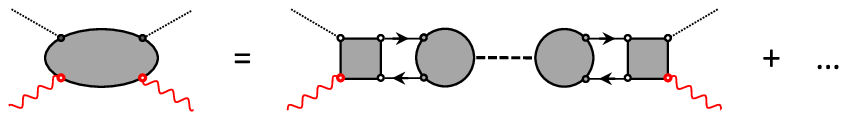}
\caption{
Four-point Green's function of photons and auxiliary scalars, $\chi$,
in the neighborhood of a pole corresponding to bound or semi-bound two-fermion state. Grey-shaded rectangles show four-point Green's functions of the  $\gamma + \chi \to e^- + p$ process.
Solid lines represent fermion propagators, and gray-shaded circles represent the Bethe-Salpeter wave functions. The dashed line shows the propagator of bound or semi-bound two-fermion state. Wavy and dotted lines, respectively, represent photons and auxiliary scalars $\chi$.
}
\label{figgauge3}
\end{center}
\end{figure}

\begin{figure} [H] %
\begin{center}
\includegraphics[angle = -90,width=0.49\textwidth]{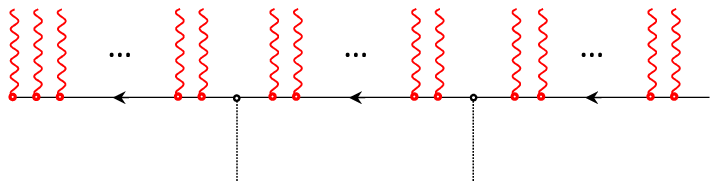}
\caption{
Graphical representation of the cut fermion loop, entering the four-point Green's function of Fig.~\ref{figgauge3}.
Wavy lines represent photons, and dashed lines represent auxiliary scalars $ \chi$.
}
\label{figgauge5}
\end{center}
\end{figure}

\section{Diagrams with auxiliary scalar}
\renewcommand{\theequation}{IV.\arabic{equation}}
\setcounter{equation}{0}

Suppose that QED+ is supplemented with an auxiliary light scalar particle, $\chi$, which leads to the $p \to e^+ + \chi$  decay with a low probability.
The effective Lagrangian can be taken as
\begin{equation}
\mathcal{L}_{\mathrm{int}} = \lambda \bar{\psi} \psi_{p}^{c}\chi + \mathrm{H.c.},
\end{equation}
where $\psi$ is the electron field, $\psi_{p}^{c}$ is the charge-conjugated proton field, and $\lambda \ll 1$. $\mathcal{L}_{\mathrm{int}}$ preserves the gauge invariance of QED+. Asymptotic states of the effective theory include photons, electrons, and $ \chi $ scalars. Proton decays into positron and $\chi$ and is treated as resonance; it no longer belongs to the set of fundamental fields and is not among asymptotic states.
The ground state of the hydrogen atom decays into the photon and the $\chi$ scalar; therefore, it is also not among asymptotic states of the extended theory. This circumstance allows using the methods developed for proving the gauge invariance of QED to QED+ supplemented with the $\chi$ scalar. It is sufficient to prove that in the extended theory, physical amplitudes are transverse in the external photon momenta and are invariant with variations in the photon propagator's longitudinal component. At the end of the proof, the scalar coupling constant $ \lambda $ will be set equal to zero.

\subsection{Hydrogen atom}

\textit{Masses of ground and excited states.}
In the considered effective theory, the bound and semi-bound states of hydrogen atom appear as resonances in the $\gamma + \chi$ channel.
A diagrammatic view of the elastic scattering of the photon on the scalar, $\chi$, is shown in Fig.~\ref{figgauge3}. In a neighborhood of the resonance, 
the scattering amplitude takes the Breit-Wigner form $\mathcal{A}(s,t) \propto 1/(s - m_H^{*2})$, where $\Re m_H^{*}$ is the 
resonance mass, $- 2\Im m_H^{*}$ is either the width $ \sim e^2 \lambda^2 $ of the ground state of a hydrogen atom, or the 
width $ \sim e^2 $ of the excited states decaying predominantly via the photon emission. The gauge invariance of $m_H^{*}$ follows 
from the gauge invariance of the diagram of Fig.~\ref{figgauge3}.
\begin{figure} [!t] %
\begin{center}
\includegraphics[angle = 270,width=0.71\textwidth]{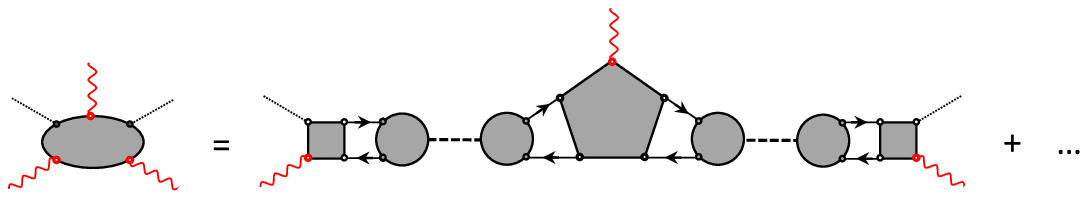}
\caption{
Five-point Green's function
in neighborhoods of poles corresponding to two-fermion bound or semi-bound states.
Notations are the same as in Figs.~\ref{figgauge1}~-~\ref{figgauge3}.
}
\label{figgauge4}
\end{center}
\end{figure}

This diagram contains a fermion loop with two or more photon vertices and two scalar vertices. The higher orders of perturbation theory for $\lambda$ can be neglected. At some point at which a photon with momentum $q$ is attached to the proton line, we cut and straighten the fermion loop, as shown in Fig.~\ref{figgauge5}. The proton part of the fermion line can also be viewed as the propagation of an antiproton. Let us consider all possible insertions of the photon vertex into the fermion line and find divergence in the photon momentum. In any gauge-invariant theory, this divergence must vanish. The antiproton propagator with the initial momentum $p$, the final momentum $p^{\prime}$ and all insertions of photon vertices is denoted by $S(p^{\prime}, p)$, and the electron propagator is $G(p^{\prime}, p)$. The WGFT identity implies that divergence of the sum of insertions of the photon vertex with momentum $q$ into the fermion propagators is equal to the fermion propagators' difference for the shifted arguments. The amplitude of Fig.~\ref{figgauge5} can be written as follows:
\begin{equation} \label{SSS}
S(p^{\prime \prime \prime}, p^{\prime \prime} + Q_{2})G(p^{\prime \prime},p^{\prime}+Q_{1})S(p^{\prime},p),
\end{equation}
where $Q_{1,2}$ are momenta transferred to the fermions in the scalar vertices, momenta transferred to the fermions in photon vertices are suppressed, the leftmost photon vertex $\hat{a}_{0}$ is omitted.
Let $\Gamma^{\mu}$ be the sum of all insertions of the photon vertex $\gamma^{\mu}$ into the amplitude of Fig.~\ref{figgauge5}. Divergence of $\Gamma^{\mu}$ consists of three terms. The first one is associated with the modification of the outer right-hand part of the fermion line (antiproton part), the second one arises from the modification of the middle part, which is the electron line between two scalar vertices,
and the third one is the result of modifying the outer left-hand part of the fermion line (antiproton part). Let $p$ be the momentum of incoming antiproton; $p^{\prime}$, its momentum before the absorption of $\chi$;
$p^{\prime} + Q_{1}$, the momentum of the electron after the absorption;
$p^{\prime \prime}$, the momentum of the electron before the absorption of $\chi$ with the momentum $Q_{2}$, and $p^{\prime \prime \prime}$, the final momentum of the outgoing antiproton.

Accordingly, one can write
\begin{eqnarray}
q_{\mu } \Gamma^{\mu } &=& S(p^{\prime \prime \prime}+q,p^{\prime \prime}+Q_{2} + q)G(p^{\prime \prime}+q,p^{\prime}+Q_{1} +q)\left( S(p^{\prime},p) - S(p^{\prime}+q,p+q)\right) \nonumber \\
                                &+&S(p^{\prime \prime \prime}+q,p^{\prime \prime}+Q_{2} + q)\left(G(p^{\prime \prime},p'+Q_{1} )-G(p^{\prime \prime}+q,p'+Q_{1} +q)\right)S(p',p) \nonumber \\
                                &+& \left(S(p^{\prime \prime \prime},p^{\prime \prime}+Q_{2} ) - S(p^{\prime \prime \prime}+q,p^{\prime \prime}+Q_{2} +q)\right)G(p^{\prime \prime},p^{\prime}+Q_{1}) S(p^{\prime},p). \label{BD}
\end{eqnarray}
In this expression, six terms occur; we enumerate them consecutively. The first term in Eq.~(\ref{BD}) cancels the fourth term. The third term cancels the sixth term. We obtain
\begin{eqnarray}
q_{\mu } \Gamma^{\mu }  &=&  -  S(p^{\prime \prime \prime}+q,p^{\prime \prime}+Q_{2} + q)G(p^{\prime \prime}+q,p^{\prime}+Q_{1} +q) S(p^{\prime}+q,p+q) \nonumber \\
 &&  + S(p^{\prime \prime \prime},p^{\prime \prime}+Q_{2} )G(p^{\prime \prime},p^{\prime}+Q_{1}) S(p^{\prime},p). \label{BD2}
\end{eqnarray}
The divergence of the closed fermion loop can be found to be
\begin{equation} \label{WGFTid}
q_{\mu } \int \frac{d^{4} p}{(2\pi )^{4} }  \mathrm{Sp}\left(\hat{a}_{0} \Gamma^{\mu } \right) = 0,
\end{equation}
where $\hat{a}_{0}$ is the leftmost photon vertex of Fig.~\ref{figgauge5}. After shift of the integration variable, $p \to p - q$,
two terms of Eq.~(\ref{BD2}) cancel each other. The divergence of the loop vanishes. This finding proves the gauge invariance of the diagram of Fig.~\ref{figgauge3}. Given that the scattering amplitude is gauge-invariant, poles corresponding to the masses of hydrogen atom are also gauge-invariant.


\textit{Electromagnetic form factors.}
The proof of the gauge invariance of the electromagnetic form factors proceeds along the line described above. Shown in Fig.~\ref{figgauge4} is Green's function with two external lines of real photons, one external line of a virtual photon, and two external lines of auxiliary scalars on the mass shell. The Green's function has simple poles corresponding to semi-bound states in the variables $s_i = P_i^2$ and $s_f = P_f^2$, where $P_i$ and $P_f$ are the total four-momenta of the $\gamma + \chi$ system in the initial and final states, respectively. We perform the Laurent series expansion of the Green's function in neighborhoods of the masses of bound or semi-bound states.
The corresponding decomposition of the Green's function is shown in Fig.~\ref{figgauge4}; the dots represent less singular contributions. The electromagnetic form factor is factorized. If the amplitude is gauge-invariant, the electromagnetic form factor is also gauge-invariant. The proof
proceeds by cutting the fermion loop as described above and verifying that all insertions of the photon vertex into the fermion line result in an expression transverse in the photon momentum. This is true both for external photon lines and for internal lines that determine the photon propagator.

\subsection{Generalized Ward--Green--Fradkin--Takahashi identity}

In Sect. IV.A, we used the generalized WGFT identity for the fermion
loops with two scalar vertices and an arbitrary number of photon vertices.
For further progress, one needs to generalize the WGFT identity for the fermion loops with arbitrary numbers of vertices of photons,
scalar $\sigma$-mesons and vector $\omega$-mesons.

Consider a fermion loop with $r+1$ vertices of the scalar and vector mesons in addition to the photon vertices. We cut it at a photon vertex $\hat{a}_0$ and straighten the fermion line. The corresponding diagram can be represented in the form similar to (\ref{SSS}):
\begin{equation}
S_{1}(\ldots)\prod\limits_{i}O_iS_{i}(\ldots)O_1S_{1}(\ldots).
\end{equation}
The index $i  = 2,\ldots,r$ denotes a fermion with either the proton mass $M$ or the electron mass $m$. The fermion propagators include the photon vertices. The momentum variables are suppressed. $O_i = 1$ and $\gamma_{\mu}$ for $i  = 1,\ldots,r$ are the scalar and vector meson vertices, respectively.
The Lorentz indices can be contracted with vectors independent of the momenta entering the vertices.
The generalized WGFT identity also applies to $O_i = \gamma_5$, $\gamma_5 \gamma_{\mu}$, $\sigma_{\mu \nu}$ for $i  = 1,\ldots,r$ and
to any combination of these vertices. It becomes possible
to supplement QED not only with protons, but also with the $\sigma$-, $\omega$-, $\pi^0$-, $\rho^0$-mesons
and any other neutral mesons involving in strong interaction.
We remark that in the classical Walecka model \cite{Walecka:1974}, nucleons exchange by the $\sigma$- and $\omega$-mesons.

Insertions of the photon vertex with the momentum $q$ into the fermion propagators result in
the divergence
\begin{eqnarray}
q_{\mu }\Gamma ^{\mu } &=&\left( -S_{1}(\ldots+q)+S_{1}(\ldots)\right)
\prod\limits_{i}O_iS_{i}(\ldots)O_1S_{1}(\ldots) \nonumber \\
&&+S_{1}(\ldots+q)\sum_{j}\prod\limits_{i<j}O_iS_{i}(\ldots+q)O_j\left(
-S_{j}(\ldots+q)+S_{j}(\ldots)\right) \prod\limits_{j<i}O_iS_{i}(\ldots)O_1S_{1}(\ldots) \nonumber \\
&&+S_{1}(\ldots+q)\prod\limits_{i}O_iS_{i}(\ldots+q)O_1\left(
-S_{1}(\ldots+q)+S_{1}(\ldots)\right),
\end{eqnarray}%
where $2 \leq j \leq r$ and $''\ldots+q''$ means that the momentum variables are shifted by $q$.
In the second line,
the first term proportional to $-S_{j}(\ldots+q)$
cancels the second term proportional to
$S_{j+1}(\ldots)$ for all $j$. The second term with the multiplier $S_{2}(\ldots)$ and the first term with the multiplier $-S_{r}(\ldots+q)$ survive and cancel the first term in the first line and the second term in the third line, respectively.
We thus obtain the identity
\begin{eqnarray}
q_{\mu }\Gamma ^{\mu } &=&-S_{1}(\ldots+q)\prod\limits_{i}O_iS_{i}(\ldots+q)O_1S_{1}(\ldots+q) \nonumber \\
&&+S_{1}(\ldots)\prod\limits_{i}O_iS_{i}(\ldots)O_1S_{1}(\ldots),
\end{eqnarray}%
which generalizes Eq.~(\ref{BD2}). Calculating the integral over the internal momentum in the loop, we arrive at Eq.~(\ref{WGFTid}),
with $\Gamma ^{\mu }$ redefined accordingly.

\begin{figure} [!t] %
\begin{center}
\includegraphics[angle = 90,width=0.3\textwidth]{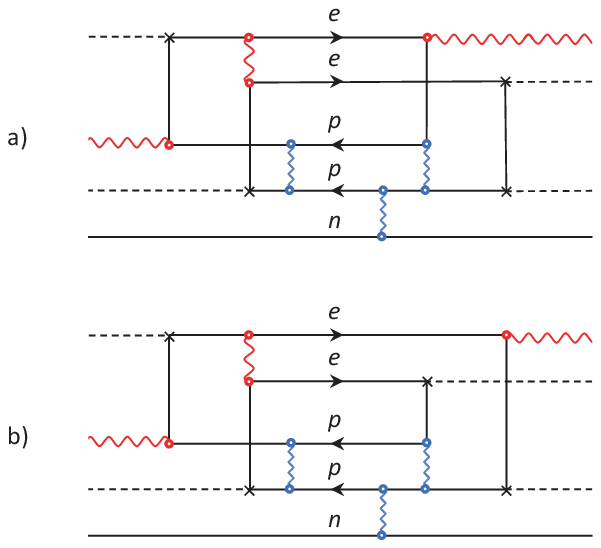}
\caption{
Skeleton diagrams of the $\gamma + 2\chi + n \to \gamma + 2\chi + n $ scattering leading to the appearance in the intermediate state of
electrically neutral atom of $^3_2$He. Diagram a) is constructed of two fermion loops, and diagram b) is constructed of a single fermion loop.
Wavy lines show photons, auxiliary scalars are shown by dashed lines, zigzag lines show the $\sigma$- and $\omega$-mesons. Arrows indicate electric charge flow.
}
\label{figgauge8}
\end{center}
\end{figure}

\subsection{Multi-electron atoms}

\textit{Masses of ground and excited states.}
In Sect.~III.B, we considered nuclei as elementary charged spin-1/2 fermions. Here nuclei are considered complex objects consisting of protons and neutrons. The binding of nucleons is provided by the exchange of $ \sigma $-, $ \omega $-, $ \pi^0 $-mesons and, possibly, other mesons, the vertices of interaction with nucleons of which do not depend on the momentum. This approach is widely used in modeling nuclei and infinite nuclear matter \cite{Walecka:1974,Oertel:2017}. In one-boson exchange models, the $ \sigma $-meson is responsible for the attractive long-range part of the nucleon-nucleon potential, while the $ \omega $-meson is responsible for the short-range repulsion. The presence of neutrons, $n$'s, among the asymptotic states is admissible due to their electroneutrality if we neglect the neutron's anomalous magnetic moment. Neutral and ionized atoms do not create asymptotic states;
therefore, the arguments used in the proofs of the gauge invariance of QED apply to the extended theory. The loops of the electrically charged fermions interacting via the meson exchange are transverse in the photon momenta, as demonstrated in Sect.~IV.B.

Figure~\ref{figgauge8} shows skeleton diagrams for the production and decay of a helium atom composed of two electrons and a nucleus with the mass and atomic numbers $A = 3$ and $Z = 2$, respectively. In the extended theory, the atom acquires a resonance status in the $\gamma + 2\chi + n$ channel. The gauge invariance of the diagrams of Fig.~\ref{figgauge8} is obvious under the generalized WGFT identity. The ground and excited states of the atom are identified as the Breit-Wigner poles of the scattering amplitude.

Combinatorial arguments show that, for any charge $Z \ge 2$,
the resonance scattering
\begin{equation}
 Z\chi + (A - Z)n \to Z\chi + (A - Z)n,
\end{equation}
via the electrically neutral atom $^A_Z$X is described by a set of skeleton diagrams
with closed fermion loops in an amount from 1 to $Z$.
These loops are connected by the $\sigma$- and $\omega$-meson propagators
without violating transverse nature in the photon momenta.
The $A - Z$  neutrons essentially play the role of spectators interacting with the protons by exchanging
the $ \sigma $-, $ \omega $-, $\ldots$ mesons.

\textit{Electromagnetic form factors.} Similar arguments apply to the process
\begin{equation}
Z\chi + (A - Z)n \to \gamma^* + Z\chi + (A - Z)n,
\end{equation}
with a virtual photon $\gamma^*$ and $Z \ge 2$.
In the channels $Z\chi + (A - Z)n$, it is necessary to localize the poles of the gauge-invariant scattering amplitude, corresponding to bound or semi-bound states. The corresponding residues
determine the gauge-invariant diagonal and transition form factors.

Setting $\lambda = 0$, we recover the quantum electrodynamics of photons, electrons, and nucleons interacting via the $ \sigma $-, $ \omega $-, $\ldots$ meson exchange.

\section{Conclusion}

The space of asymptotic states of the quantum electrodynamics of photons, electrons, and nuclei
contain multi-electron atoms as bound states. This circumstance does not allow the direct extension of standard proofs of the gauge invariance of QED to processes involving atoms.
We described two ways to exclude atoms from the space of asymptotic states:

The first is associating each scattering process with incoming and outgoing atoms to another scattering process with the same number of electrons and the same nuclei
above the elastic threshold. By construction, the incoming and outgoing asymptotic states of the associated process do not contain bound states; therefore, the standard proofs of QED's gauge invariance apply to it. The process amplitude contains poles corresponding to bound and semi-bound states when continued analytically in invariant masses of the appropriate sub-channels. The residues of the amplitude emerge as the gauge-invariant quantities; they determine properties of multi-electron atoms. We showed how exactly the masses of bound and semi-bound states determine the forward scattering amplitude's analytical properties and the spectral support regions of the  $ e^- + p \to e^- + p $ process.

The second method extends QED by adding to the theory nucleons, neutral mesons, and an auxiliary scalar particle, $\chi$, which causes the proton decay $p \to e^+ + \chi$ with a low probability. The asymptotic states are formed by photons, electrons, neutrons, and $\chi$-scalars, while multi-electron atoms emerge as resonances.
The standard proofs of the gauge invariance of QED apply to the scattering amplitudes of the extended theory. The gauge invariance pertains to the masses and electromagnetic form factors of multi-electron atoms. We treated nuclei as collections of nucleons bound by strong forces mediated by neutral mesons ($\sigma$, $\omega$, $\ldots$).
A generalization of the WGFT identity to the quantum electrodynamics of photons, electrons, and nucleons, interacting with electrically neutral mesons, and
$\chi$-scalars, was used to demonstrate the transverse nature of scattering amplitudes of the extended theory. After completing the proof, the coupling constant of $\chi$ is set equal to zero to restore the initial, physical set of particles.

We demonstrated the gauge invariance of the masses and electromagnetic form factors of multi-electron atoms. The proposed methods can be generalized to the scattering amplitudes with arbitrary numbers of photons, electrons, nuclei, and neutral and ionized atoms in the ground state. The results presented indicate that multi-electron atoms are, in general, gauge-invariant objects.
Ambiguities in various models' predictions are related exclusively to the uncertainties inherent in the atomic structure calculations. Concurrently, in high orders of perturbation theory, the use of any gauge is justified.

\acknowledgments

This work was supported in part by the RFBR Grant No.~18-02-00733.

\end{document}